\documentstyle[aps,prl,epsfig]{revtex}
\draft
\newcommand{\size}{6.5}

\begin{document}
\twocolumn[\hsize\textwidth\columnwidth\hsize\csname@twocolumnfalse\endcsname
\title{Pauli blocking factors in quantum pumps}
\author{Sang Wook Kim}
\address{Max-Planck-Institut f{\"u}r Physik komplexer Systeme,
N{\"o}thnitzer Str. 38, D-01187 Dresden, Germany}
\date{\today}

\maketitle

\begin{abstract}

We investigate the Pauli blocking factor in quantum pumps using Floquet formalism. 
Even though the time dependent potentials in quantum pumping can not only cause 
inelastic scatterings but also break the micro-reversibility, i.e. 
$T^+(E',E) \neq T^-(E,E')$, the Pauli blocking factor is unnecessary when 
the scattering process through the scatterer is coherent. The well defined scattering 
states extending from one reservoir to the others form a complete non-orthogonal set. 
Regardless of the non-orthogonality one can obtain the pumped currents using the 
field operator formalism. The current expression finally obtained do not contain 
Pauli blocking factor.

\end{abstract}

\narrowtext
\vskip1pc]


Quantum transport through artificially fabricated nano/mesostructures has been 
extensively studied both experimentally and theoretically during the past years. 
Conductances between two reservoirs through the nano/mesostructures can be calculated 
by using Landauer-B\"uttiker formalism \cite{Datta95} when no dephasing process occurs
in the scatterer region. The current through the scatterer can then be obtained from
\begin{equation}
I = \frac{e}{h}\int dE dE' \left[ T^+(E',E)f_L(E) - T^-(E,E')f_R(E') \right],
\label{no pauli current}
\end{equation}
where $T^+(E',E)$ represents the transmission probability for scattering states incident 
from the left at energy $E$ and emerging to the right at $E'$, and $T^-(E,E')$ is defined 
in a similar manner for the reverse direction. $f_L$ ($f_R$) is the Fermi-Dirac 
distribution in the left (right) reservoir. There has been some debate and 
confusion for using this formula 
\cite{Datta95,Hekking91,Landauer92,Datta92a,Datta92b,Sols92,Boenig93,Wagner00}
since Eq.~(\ref{no pauli current}) dose not contain the so-called Pauli blocking factors.
The fermionic nature of the electrons is taken care of in an {\it ad hoc} way by factors
$1-f$ to suppress scattering into occupied states, so that the current is given by
\begin{eqnarray}
I & = & \frac{e}{h}\int dE dE' \left\{ T^+(E',E)f_L(E) \left[ 1-f_R(E') \right] \right. \nonumber \\
  & - & \left. T^-(E,E')f_R(E') \left[ 1-f_L(E) \right] \right\}.
\label{pauli current}
\end{eqnarray}
Usually these two expressions Eqs.~ (\ref{no pauli current}) and (\ref{pauli current}) give 
the same results since the difference between them,
$[T^+(E',E)-T^-(E,E')]f^L(E)f^R(E')$, vanishes when $T^+(E',E)=T^-(E,E')$, i.e., the 
micro-reversibility holds. The question can arise, however, if the system lack of this
micro-reversal symmetry is considered. One of the relevant example is a quantum pump.

The quantum pump is a device that generates a dc current at zero bias potential through
cyclic change of system parameters \cite{Altshuler99,Lubkin99}. Recently, adiabatic charge pumping 
in open quantum dots has attracted considerable attention 
\cite{Aleiner98,Brouwer98,Zhou99,Wei00,Avron00,Shutenko00,Avron01,Sharma01,Entin-Wohlman02,Moskalets02a},
and was experimentally realized by Switkes et al. \cite{Switkes99}.
After a cycle of the adiabatic shape change we return to the initial configuration, but 
the wavefunction may have its phase changed from the initial wavefunction. This is the 
geometric or Berry's phase \cite{Berry84}. The additional phase is equivalent 
to some charges that pass through the quantum dot, namely, pumped charge \cite{Altshuler99}. 
In another point of view, the quantum pump is a time dependent system driven by (at least) 
two different time periodic perturbations with the same angular frequency and a phase 
difference. One can deal with this problem using not only adiabatic approximation 
but also Floquet approach \cite{Sambe73,Kim02_qp,Moskalets02b}. Recently, it has been shown that 
in the adiabatic limit with small strength of the oscillation potentials the Floquet and the 
adiabatic approach give exactly equivalent results \cite{Kim02_qp}.

More than two periodically oscillating perturbations with a phase difference
($\neq n\pi$, $n$ is an integer) break the time reversal symmetry \cite{Wagner00}, and consequently
$T^+(E',E) \neq T^-(E,E')$. Therefore, the currents obtained from Eq.~(\ref{no pauli current}) 
and (\ref{pauli current}) are different from each other in quantum pumps 
\cite{Datta92b,Wagner00}. The question immediately arises: 
which one is correct in quantum pumps? It is noted that this problem still exists even in 
the adiabatic limit, which will be shown below. We would like to make a conclusion first. 
Even though the time dependent scatterer like the case of quantum pumping can not only cause 
inelastic scatterings but also break the mico-reversibility, 
{\em the Pauli blocking factor is unnecessary} when the scattering process through 
the scatterer is {\em coherent}. 

The existence of the Pauli blocking factors is intimately related to the ``scattering
states'' \cite{Datta95}. If we fill up the energy eigenstates with the electrons in both 
reservoirs independently and then transfer the electrons  from one to the other reservoir, 
the Pauli blocking factors cannot be unavoided. If the transport is coherent across the 
scatterer, however, one can define a single wavefunction extending from one reservoir to the 
other (more precisely reflected and transmitted waves in every connected reservoirs) and 
then fill up these scattering states. In this consideration the concept of transferring 
the electron from one to the other reservoir is automatically eliminated, so is the Pauli 
blocking. The scattering states of the problem with static scatterer was proven to be 
orthogonal and complete \cite{Kriman87}. 

Consider the Schr\"odinger equation $i\hbar(\partial/\partial t)\psi = H(t)\psi$
for an elctron with mass $\mu$ and $H(t) = -\nabla^2/2\mu + U(x,t)$, where $U(x,t+T)=U(x,t)$ 
and $U(x,t)=0$ at $x\rightarrow \pm\infty$. For an energy $E=\hbar^2 k^2/2\mu$ ($k>0$) 
of the incoming particle the scattering states as a solution of the Schr\"odinger equation 
can be defined as
\begin{eqnarray}
&&\chi^+_E(x,t) = \nonumber \\
 	&&\left\{ 
		\begin{array}{ll}
			e^{ikx-iEt/\hbar} + \sum_{E_n>0} r^+_{E_nE}
				e^{-ik_nx-iE_nt/\hbar}, & x \rightarrow -\infty, \\
			\sum_{E_n>0} t^+_{E_nE} e^{ik_nx-iE_nt/\hbar}, 
				& x \rightarrow +\infty, 
		\end{array}
	\right.
\label{scatt1}
\\
&&\chi^-_E(x,t) = \nonumber \\
	&&\left\{
		\begin{array}{ll}
			e^{-ikx-iEt/\hbar} + \sum_{E_n>0} r^-_{E_nE}
				e^{ik_nx-iE_nt/\hbar}, & x \rightarrow +\infty, \\
			\sum_{E_n>0} t^-_{E_nE} e^{-ik_nx-iE_nt/\hbar}, 
				& x \rightarrow -\infty, 
		\end{array}
	\right.
\label{scatt2}
\end{eqnarray}
where $E_n = E+n\hbar\omega$, $k_n=\sqrt{2\mu E_n}/\hbar$, and the normalization is ignored. 
Here we have reflection and transmission coefficients $r^+_{E_nE}$ and $t^+_{E_nE}$, 
which can be obtained from {\em unitary} Floquet scattering matrices $S$ 
\cite{Kim02_qp,Henseler01,Kim02_pole}. The matrix $S$ has the following form
\begin{equation}
S(\epsilon) = \left(
        \begin{array}{cccccc}
        {\bf r}_{00} & {\bf r}_{01} & \cdots & {\bf t'}_{00} & {\bf t'}_{01} & \cdots \\
        {\bf r}_{10} & {\bf r}_{11} & \cdots & {\bf t'}_{10} & {\bf t'}_{11} & \cdots \\
        \vdots & \vdots & \ddots & \vdots  & \vdots  & \ddots \\
        {\bf t}_{00} & {\bf t}_{01} & \cdots & {\bf r'}_{00} & {\bf r'}_{01} & \cdots \\
        {\bf t}_{10} & {\bf t}_{11} & \cdots & {\bf r'}_{10} & {\bf r'}_{11} & \cdots \\
        \vdots & \vdots & \ddots & \vdots  & \vdots  & \ddots \\
        \end{array}
\right),
\label{osc_smatrix}
\end{equation}
where ${\bf r}_{\alpha\beta}$ and ${\bf t}_{\alpha\beta}$ are the reflection and the 
transmission amplitudes respectively, for modes incident from the left with an Floquet energy 
$\epsilon$ which take continuous values in the interval $[0,\hbar \omega)$; 
${\bf r'}_{\alpha\beta}$ and ${\bf t'}_{\alpha\beta}$ are similar quantities for modes 
incident from the right. The above transmission and reflection coefficients are related to 
the matrix elements of $S$ in terms of 
$t^+_{E_n,E} = \sqrt{k_\beta/k_\alpha}{\bf t}_{\alpha\beta}$, {\it etc.} with 
$E=\epsilon+\beta\hbar\omega$ and $\alpha=n+\beta$.

For $E'-E\neq m\hbar\omega$ ($m$ is an integer) the orthogonality of these scattering states 
can be immediately proven by using the orthogonality of the functions $e^{\pm ik}$. 
When $E'-E= m\hbar\omega$, however, we obtain at any fixed time
\begin{eqnarray}
&&\left< \chi^+_E \left|\right. \chi^-_{E'}\right>  \nonumber \\
 & \propto & \sum_{E_n>0} \left( r^{+*}_{E_n,E} t^-_{E_n,E+m\hbar\omega}
 + t^{+*}_{E_n,E}r^-_{E_n,E+m\hbar\omega} \right) \nonumber \\
 & = & \sum_{\alpha} \frac{\sqrt{k_\beta k_{\beta+m}}}{k_\alpha}
 \left( {\bf r^*}_{\alpha\beta} {\bf t'}_{\alpha,\beta+m}
 + {\bf t^*}_{\alpha\beta} {\bf r'}_{\alpha,\beta+m} \right) \label{ortho1}\\
 & \propto & \sum_{\alpha} s^*_{\alpha\beta}s_{\alpha,\beta+m}/k_{\alpha},
\nonumber
\end{eqnarray}
and
\begin{eqnarray}
&&\left< \chi^+_E \left|\right. \chi^+_{E'}\right> \nonumber \\ 
 & \propto & \delta_{E,E'} + \sum_{E_n>0} \left( r^{+*}_{E_n,E}r^+_{E_n,E+m\hbar\omega}
 + t^{+*}_{E_n,E}t^+_{E_n,E+m\hbar\omega} \right) \nonumber \\
 & = & \delta_{\beta,\beta+m} + 
       \sum_{\alpha} \frac{\sqrt{k_\beta k_{\beta+m}}}{k_{\alpha}}
       \left( {\bf r^*}_{\alpha\beta} {\bf r}_{\alpha,\beta+m} 
       + {\bf t^*}_{\alpha\beta} {\bf t}_{\alpha,\beta+m} \right) \label{ortho2}\\
 & = & \delta_{m,0} + \sqrt{k_\beta k_{\beta+m}}
       \sum_{\alpha} s^*_{\alpha,\beta}s_{\alpha,\beta+m}/k_{\alpha}. 
\nonumber
\end{eqnarray}
One can find the same result for $\left< \chi^-_E \left|\right. \chi^-_{E'}\right>$
using the similar procedure used in Eq.~(\ref{ortho2}).
The unitarity of the Floquet scattering matrix does not guarantee the orthogonality of 
the scattering states. In the multichannel scattering problem with a static scatterer
the orthogonality is drawn from the orthogonality of the channel eigenfunctions.

We consider the completeness of the scattering states. A solution of the time periodic 
Hamiltonian can be formally written as
\begin{equation}
\Psi_\epsilon(x,t) = e^{-i \epsilon t/\hbar}\sum_{n=-\infty}^{\infty} \psi_n (x) e^{-in\omega t}.
\label{floquet}
\end{equation}
Since the potential is zero at $x\rightarrow \pm \infty$, $\psi_n(x)$ is given by the
following form
\begin{equation}
\psi_n (x) = \left\{
\begin{array}{c}
A_n e^{ik_n x} + B_n e^{-ik_n x}, ~~~ x\rightarrow -\infty \\
C_n e^{ik_n x} + D_n e^{-ik_n x}, ~~~ x\rightarrow +\infty,
\end{array} \right.
\label{plane_wave}
\end{equation}
where $k_n=\sqrt{2\mu(\epsilon + n\hbar\omega)}/\hbar$. One can immediately know that the 
linear combination of $A_n \chi^+_E + D_n \chi^-_E $ completely cover all Floquet 
type solutions. The scattering states form a complete set for describing the solution of
a scattering problem with time periodic potential. We have shown the scattering states 
$\chi^\sigma_E$ form a complete non-orthogonal set.

We derive the current using these scattering states. The time dependent electron field 
operator can be obtained in the following form \cite{Buettiker92,Levinson99}
\begin{equation}
\Psi(x,t) = \sum_\sigma \int dE ~ \chi^\sigma_E(x,t) \frac{a_{\sigma E}}{\sqrt{hv(E)}},
\label{expansion}
\end{equation}
where $a_{\sigma E}$ and $v(E)$ is an annihilation operator for electrons in the scattering 
states $\chi^\sigma_E(x,t)$ and the velocity, respectively. Even though the scattering states 
do not form othogonal bases we need only the completeness to be sure that the expansion of 
Eq.~(\ref{expansion}) is valid. Using this field operator the current operator 
is also expressed as
\begin{eqnarray}
J(x,t) & = & (ie/2m)\Psi^+(x,t)\nabla \Psi(x,t) + H.c. \nonumber \\
 & = & \frac{e}{m}\sum_{\sigma\sigma'} \int dE dE'
       ~ {\rm Im}(\chi^{\sigma' *}_{E'} \nabla \chi^{\sigma}_E)
        \frac{a^+_{\sigma E} a_{\sigma' E}} {h\sqrt{v(E)v(E')}}.
\end{eqnarray}

\begin{figure}
\center
\includegraphics[height=\size cm,angle=0]{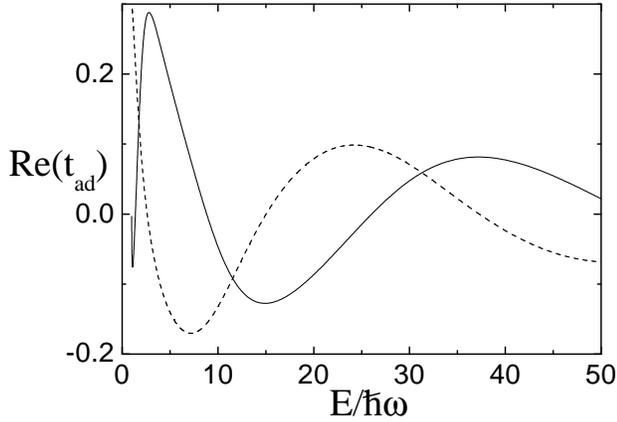}
\caption{${\rm Re}[t_{ad,1}(E)]$ (the solid curve) and ${\rm Re}[t_{ad,-1}(E+\hbar\omega)]$ 
with $\lambda = 22.5$ meV$\cdot$nm, $\phi=\pi/2$, $\mu=0.067m_e$, $d=50$ nm, and $T=9.09$ ps.}
\label{fig1}
\end{figure}

The quantum mechanical (or thermal) average of the current operator becomes
\begin{eqnarray}
\left< J(x,t) \right> 
 &=& \frac{e}{m}\sum_\sigma \int dE ~ {\rm Im}(\chi^{\sigma *}_{E} \nabla \chi^{\sigma}_E) 
   \frac{\left< a_{\sigma E}^+a_{\sigma E} \right>}{hv(E)}  \nonumber \\
 &+& \frac{e}{m}\sum_{\sigma \sigma'} \sum_{E_n>0, n \neq 0} \int dE \label{current_avg} \\
 &\times&     {\rm Im}(\chi^{\sigma' *}_{E_n} \nabla \chi^{\sigma}_E)
      \frac{\left< a_{\sigma' E_n}^+a_{\sigma E} \right>}{h\sqrt{v(E)v(E_n)}}. 
\nonumber
\end{eqnarray}
We evaluate Eq.~(\ref{current_avg}) taking $x \rightarrow \infty$ and averaging over 
space and time. One can then obtain the pumped current as following
\begin{equation}
I = \frac{e}{h}\sum_{E_n>0}\int dE \left[T^+_{E_nE}f_L(E) - T^-_{E_nE}f_R(E)\right],
\label{final}
\end{equation}
where we exploit $\sum_{E_n} (k_n/k)|r^-_{E_nE}|^2= 1-\sum_{E_n} (k_n/k)|t^-_{E_nE}|^2$ and 
$\left< a_{\sigma E}^+a_{\sigma E} \right> = f_\sigma(E)$. The second term of the righthand 
side in Eq.~(\ref{current_avg}) vanishes due to the unitarity of the scattering matrix
\cite{cal_detail}. Here, $T^{\pm}_{E_nE}$ denotes $(k_n/k)|t^{\pm}_{E_nE}|^2$. The final 
result, Eq.~(\ref{final}), exactly corresponds to Eq.~(\ref{no pauli current}), i.e. 
the current {\em without} Pauli blocking factor.

Now we consider the adiabatic limit. The adiabatic condition in the quantum pump implies 
that any time scale of the problem considered must be much smaller than the period of the 
oscillation of an external pumping \cite{Brouwer98}. We can then define the instantaneous 
scattering matrix with time dependent parameters, namely $X_n(t)$,
\begin{equation}
S_{ad}(E,t) = S_{ad}(E,X_1(t), X_2(t), \cdots).
\end{equation}
Due to the time periodicity of $X_n$'s, using Fourier transform one can obtain the amplitudes
of side bands for 
\begin{figure}
\center
\includegraphics[height=\size cm,angle=0]{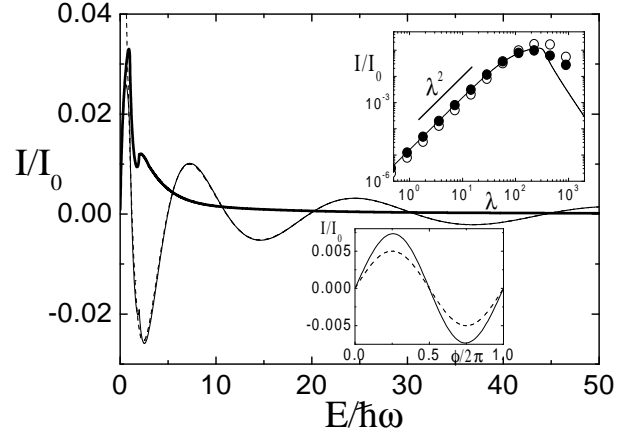}
\caption{The pumped currents obtained from the Floquet approach with (the thin solid
curve) and without Pauli blocking (the thick solid curve), and Brouwer's formula 
(the dashed curve) with the same parameters used in Fig.~\ref{fig1}.
Note that $I_0=e/T=17.6$ nA.
The upper inset: the pumped currents as a function of the strength of the oscillating
potential $\lambda$ from the Floquet approach with (the filled circles) and without Pauli 
blocking (the open circles), and Brouwer's formula (the solid curve). The lower inset:
the pumped currents as a function of a phase difference $\phi$ from the Floquet approach 
with (the solid curve) and without Pauli blocking (the dashed curve). In both insets
we take $E=6.005\hbar\omega$}
\label{fig2}
\end{figure}
particles traversing the adiabatically oscillating scatterer with incident
energy $E$ as following \cite{Moskalets02b}
\begin{equation}
S_{ad,n}(E)=\frac{1}{T}\int_0^T dt~e^{in\omega t}S_{ad}(E,X_1(t), X_2(t), \cdots).
\end{equation}
The micro-reversibility condition in this expression is given by 
$t_{ad,n}(E)=t'_{ad,-n}(E+n\hbar\omega)$, where $t_{ad}$ and $t'_{ad}$ represent the 
adiabatic transmission amplitude for the forward and the backward direction, respectively. 

We show, in Fig.~\ref{fig1}, that $t_{ad,1}(E)$ and $t_{ad,-1}(E+\hbar\omega)$ for a simple 
model system, a 1D two harmonically oscillating $\delta$-function barriers with the 
strengthes $X_1=\lambda \cos \omega \tau$ and $X_2=\lambda \cos (\omega \tau + \phi)$
respectively, separated by a distance $d$ \cite{Wei00}, are clearly deviated from each 
other, which implies the micro-reversibility of adiabatic quantum pumps is also broken. 
Note that in this model system $t_{ad}(E)=t'_{ad}(E)$. The Pauli blocking 
factor was ignored in Brouwer's approach since Brouwer's theory is based upon a formula 
due to B\"uttiker, Thomas, and Pr\^etre \cite{Buettiker94}, where they obtained the current 
operator from the difference between the incoming and the outgoing distributions of the 
electrons without Pauli blocking factors. This current can also be acquied by using the 
scattering states, consequently without Pauli blocking factor \cite{Buettiker92}.

Figure \ref{fig2} shows the pumped currents obtained from Floquet approach with and 
without Pauli blocking factor, and Brouwer's formula in the same model used above
under the adiabatic regime (the Wigner delay time is much smaller than $T$ \cite{Kim02_qp})
and with small amplitudes of the oscillating strength. It is clearly
seen that the current with Pauli blocking factor deviates from that of Brouwer's approach
which nearly coincide with the current without Pauli blocking factor \cite{Kim02_qp}.
It is worth nothing that qualitative behavior of the pumped currents with and without
Pauli blocking looks quite similar: the pumped current $I \propto \lambda^2\sin\phi$
as shown in the insets of Fig.~\ref{fig2}. In Ref. \cite{Wagner00}, however, 
the temperature dependence of the pumped currents was expected to be distinct.

Finally, it is noted that if any kind of dephasing such as electron-electron interaction
and electron-phonon interaction is involved in the scattering processes it is still an open 
problem whether the Pauli blocking factor is required.

In conclusion, we have shown that the Pauli blocking factor is unnecessary in quantum pumps
when the scattering process through the quantum pump is coherent.
The well defined scattering states form a complete non-orthogonal set. One can obtain 
the pumped currents without the Pauli blocking factor using the field operator formalism
with these scattering states. Even in the adiabatic limit the problem of Pauli blocking
factor still exists. The Pauli blocking factor was ignored in Brouwer's adiabatic formalism, 
so that in the adiabatic limit with small strength of the pumping potential the pumped currents 
obtained from the Floquet theory without Pauli blocking factor show good agreement with those 
drawn from Brouwer's formula.


I would like to thank Henning Schomerus, Mikhail Titov, and Hwa-Kyun Park for 
useful discussions, and G. Cuniberti for careful reading of my manuscript.


\bibliographystyle{prsty}

\begin{thebibliography}{10}
\bibitem{cal_detail}
Due to $\nabla$, we obtain the same formula as Eqs.~(\ref{ortho1}) 
and (\ref{ortho2}) without $k_\alpha$ at the denominators.  One can easily find that 
$\overline{\chi^{\sigma'*}_{E'} \nabla \chi^\sigma_E} \propto \delta_{\sigma'\sigma} 
\delta(E-E')$, where $\overline{(\cdots)}$ represents space and time average at
$x\rightarrow +\infty$.
\end{thebibliography}

\end{document}